\documentstyle[epsf]{article}

\newcommand{\simlt} 
{\mbox{\raisebox{-0.5ex}{$ \; \sim$}
\raisebox{0.8ex}{$ \!\!\!\!\!\!\! <$}}}
\begin{document}
\mbox{ }
\rightline{UCT-TP-262/04}\\
\rightline{September 2004}\\
\vspace{3.5cm}
\begin{center}
{\Large \bf Electromagnetic  proton form factors in large $N_{c}$ 
 QCD}\footnote{Work supported
in part by the Volkswagen Foundation}\\
\vspace{.5cm}
{\bf C. A. Dominguez, T. Thapedi}\\[.5cm]
Institute of Theoretical Physics and Astrophysics\\
University of Cape Town, Rondebosch 7700, South Africa\\[.5cm]
\end{center}
\vspace{.5cm}
\begin{abstract}
\noindent
The electromagnetic form factors of the proton are obtained using a 
particular realization of QCD in the large $N_c$ limit 
($\mbox{QCD}_{\infty}$), which sums up the infinite number of zero-width
resonances to yield an Euler's Beta function (Dual-$\mbox{QCD}_{\infty}$).
The form factors $F_1(q^2)$ and $F_2(q^2)$, as well as $G_M(q^2)$
agree very well  with reanalyzed space-like data in the whole range of
momentum transfer. In addition, the predicted ratio $\mu_p G_E/G_M$
is in good agreement with  recent polarization transfer measurements
at Jefferson Lab.

\end{abstract}
\newpage
\setlength{\baselineskip}{1.5\baselineskip}
\noindent

Quantum Chromodynamics (QCD) in the limit of large number of colours
($\mbox{QCD}_{\infty}$) \cite{GTH} is known to predict a hadronic
spectrum consisting of an infinite number  of zero-width resonances\cite{W}.
However, since real QCD has never been solved exactly and analytically,
the hadronic parameters  (masses, couplings, etc.) remain unpredicted.
A few models of
this spectrum have been proposed for heavy quark Green's functions 
\cite{SH}-\cite{CAD1}, as well as for light quark systems \cite{DR}.
The infinite number of zero-width resonances of $\mbox{QCD}_{\infty}$ is
reminiscent of Veneziano's dual-resonance
model \cite{VEN}, the precursor of string theory.
In fact, inspiration from this model has led to a proposal called
Dual-$\mbox{QCD}_{\infty}$ \cite{CADPI}, a specific realization
of $\mbox{QCD}_{\infty}$ where the masses and couplings in a Green's
function are  chosen to yield an Euler's  Beta function of the Veneziano
type. For three-point functions, the form factors exhibit asymptotic
Regge-behaviour, i.e. power-behaviour, in the space-like region controlled
by a single free parameter. Dual-$\mbox{QCD}_{\infty}$ has been applied
quite successfully to the electromagnetic form factor of the pion in the
space-like region \cite{CADPI}. Indeed, results are in excellent agreement
with experiment, far better than e.g. naive Vector Meson (rho-) Dominance or
purely perturbative QCD \cite{CADPION}.
This is  the case not only for the pion form factor itself, but also for
the mean-square radius, and the observed deviation from universality
(the  ratio $g_{\rho \pi \pi}/f_{\rho}$) . In addition, unitarization
of Dual-$\mbox{QCD}_{\infty}$ leads to a prediction of the vector
two-point  spectral function, in the time-like region, in reasonable
agreement with  data (a more refined model in the time-like region has been
proposed recently in \cite{BRUCH}). Encouraged by this success,
we discuss in this note an analysis of the electromagnetic proton form
factors in the framework of Dual-$\mbox{QCD}_{\infty}$.\\

The Dirac and Pauli form factors of the proton, $F_1(q^2)$ and $F_2(q^2)$,
respectively, are defined as
%Eq.1
\begin{equation}
< N(p_{2}) | J_{\mu}^{EM}(0) | N(p_{1})> = \bar{u}_{N} (p_2)\; [ F_1(q^2)
\gamma_{\mu} + \frac{i \kappa}{2 M_p} F_2(q^2) \sigma_{\mu \nu} q^{\nu} ]\;
u_{N}(p_1) \; ,
\end{equation}
where $q^{2}=(p_{2}-p_{1})^{2}$, while $\kappa \equiv \mu_p -1$, and $M_p$ are 
the proton's magnetic moment and mass, respectively, and $F_{1,2}(q^2)$
are normalized as $F_1(0) = F_2(0) =1$. On the other hand, the Sachs' form
factors $G_E(q^2)$ and $G_M(q^2)$ are  given by
%Eq.2
\begin{eqnarray}
\begin{array}{lcl}
G_E(q^2) &=& F_1(q^2) - \kappa \;\tau \; F_2(q^2)  \\
G_M(q^2) &=& F_1(q^2) + \kappa \; F_2(q^2) \; ,
\end{array}
\end{eqnarray}

where $\tau = - q^2/4M_p^2 \equiv Q^2/4M_p^2$, and the normalization is then
$G_E(0) = 1$, and $G_M(0) = \mu_p$.\\
In the very early applications of the dual-resonance model to three point
functions involving more than one form factor \cite{EARLY}, it was not quite
clear to which form factor, or linear combination of form factors, should
the model apply. There is no ambiguity in Dual-$\mbox{QCD}_{\infty}$,
as this is a realization of a quantum field theory. The form factors should
then be those appearing in the primary hadronic spectral function, dual to
the QCD field theory spectral function. In other words, the form factors
with the correct pole structure satisfying dispersion relations in the
complex energy plane. In the case of the nucleon, these are the Dirac
and Pauli form factors which in $\mbox{QCD}_{\infty}$ become
%Eq.3
\begin{equation}
F_{1,2}(s) = \sum_{n=0}^{\infty}
\frac{C_{(1,2)n}}{(M_{n}^{2} -s)} \; ,
\end{equation}
where $s \equiv q^{2}$, and the masses of the vector-meson zero-width
resonances, $M_n$, as well as their couplings $C_{1n}$ and $C_{2n}$,
are not predicted. In Dual-$\mbox{QCD}_{\infty}$ these are chosen so that
the form factors become Beta functions (ratios of gamma-functions), i.e.
%Eq.4
\begin{equation}
C_{(1,2)n} = \frac{\Gamma(\beta_{1,2}-1/2)}{\alpha' \sqrt{\pi}} \;
\frac{(-1)^n} {\Gamma(n+1)} \;
\frac{1}{\Gamma(\beta_{1,2}-1-n)} \; ,
\end{equation}
where $\beta_{1,2}$ are free parameters controlling the asymptotic behaviour
in the space-like region ($s<0$), and $\alpha' = 1/2 M_{\rho}{^2}$ is
the universal string tension in the rho-meson Regge trajectory
%Eq.5
\begin{equation}
\alpha_{\rho}(s) = 1 + \alpha ' (s-M_{\rho}^{2}) \; .
\end{equation}
The mass spectrum is chosen as \cite{AB}
%Eq.6
\begin{equation}
M_{n}^{2} = M_{\rho}^{2} (1 + 2 n) \; .
\end{equation}
Using Eqs.(4) and (6) in Eq.(3) one obtains
%Eq.7
\begin{eqnarray}
F_{1,2}(s) &=& \frac{\Gamma(\beta_{1,2}-1/2)}{\sqrt{\pi}} \;
\sum_{n=0}^{\infty}\;
\frac{(-1)^{n}}{\Gamma(n+1)} \; \frac{1}{\Gamma(\beta_{1,2}-1-n)} \; 
\frac{1}
{[n+1-\alpha_\rho(s)]} \nonumber \\ [.2cm]
& = &
\frac{1}{\sqrt{\pi}} \; \frac{\Gamma (\beta_{1,2}-1/2)}{\Gamma
(\beta_{1,2}-1)} \;\;
B(\beta_{1,2} - 1,\; 1/2 - \alpha' s) \;,
\end{eqnarray}
where B(x,y) is Euler's Beta function. In the time-like region ($s>0$) the
poles of the Beta function  correspond to an
infinite set of zero-width resonances with equally spaced squared masses
given by Eq.(5). In fact, from Eq.(7) it follows
%Eq.8
\begin{equation}
Im \; F_{1,2}(s) = \frac{\Gamma(\beta_{1,2}-1/2)}{\alpha' \sqrt{\pi}} \;
\sum_{n=0}^{\infty} \; \frac{(-1)^{n}}{\Gamma(n+1)} \;
 \frac{1}{\Gamma(\beta_{1,2}-1-n)} \; \pi \; \delta(M_n^2-s) \;.
\end{equation}
Asymptotically, the form factors in the space-like region exhibit
Regge-behaviour, viz.
%Eq.9
\begin{equation}
\lim_{s \rightarrow - \infty} F_{1,2}(s) = (- \alpha' \;s)^{(1-\beta_{1,2})}  \; ,
\end{equation}
The free parameters $\beta_{1,2}$ can be fixed from  fits to the data in
the space-like region. Notice that the values $\beta_{1,2} = 2$ reduce the
form factors to  single rho-meson dominance (naive Vector Meson Dominance).
The mass formula Eq.(6) predicts e.g. for the first three radial excitations:
$M_{\rho'} \simeq 1340$ MeV, $M_{\rho''} \simeq 1720$ MeV,
and $M_{\rho'''} \simeq 2034$ MeV
in reasonable agreement with experiment \cite{PDG} : $M_{\rho'} = 1465 \pm
25$ MeV, $M_{\rho''} = 1700 \pm 20$ MeV,
and $M_{\rho'''} = 2149 \pm 17$ 
MeV. Alternative (non-linear) mass formulas might be required if one were to
match the asymptotic Regge behaviour to the Operator Product Expansion of
current correlators at short distances \cite{ESPRIU}. However, the
differences in the values of the first few masses are at the level of a few
percent. Hence, the form factors would hardly be affected, since the
contribution from high mass states is factorially suppressed.\\

Historically, the Sachs form factors were first determined from measurements
of elastic electron-proton scattering cross sections (Rosenbluth technique)
\cite{ROS}.
Direct extractions of $G_E(q^2)$ and $G_M(q^2)$ up to $- q^2 \equiv Q^2
\simeq 7 \; \mbox{GeV}^2$ indicated the empirical approximate scaling 
relation:
$\mu_p G_E(q^2)/G_M(q^2) \simeq 1$. At higher values of $Q^2$,  the
contribution of $G_E(q^2)$ to the cross section is kinematically suppressed.
On the other hand, recent electron-proton polarization transfer measurements
at Jefferson Lab (JLab) \cite{RATIO2}
up to $Q^2 \simeq 6 \; \mbox{GeV}^2$ show a considerable deviation from
this scaling behaviour, except possibly at very small $Q^2$  \cite{RATIO1}.
After some debate about the source of the discrepancy
between cross-section (Rosenbluth) and polarization transfer extractions of
the form factors \cite{ARR1}, it appears that the culprit is the two-photon
exchange correction \cite{2GAMMA}. We  assume
this to be the case, and adopt
the experimental data on the Sachs form factors as corrected in \cite{BRASH}.
These corrections are made in order to bring the Rosenbluth data into
agreement  with the polarization transfer data on the ratio
$\mu_p G_E(q^2)/G_M(q^2) $.
We then  use Eq.(2) to obtain  {\it data points} for $F_1(q^2)$ and
$F_2(q^2)$. After fitting this data base with Eq.(7) we find
$\beta_1 = 3.03$, and $\beta_2 = 4.20$.
Figures 1, and 2 show the results of the fits for $F_1(q^2)$, and $F_2(q^2)$,
corresponding to these values of $\beta_{1,2}$,
together with the corrected data points of \cite{BRASH}. In Fig. 3 we
show $G_M(q^2)$, as obtained from Eq.(2) using the fitted $F_{1,2}(q^2)$,
together with the same data base. As can be appreciated, the agreement
between Eq.(7) and the data is very good. Having fitted a data base
corrected to account for the polarization transfer data on the ratio
$\mu_p G_E(q^2)/G_M(q^2) $, we would expect our theoretical form factors
to lead to a ratio in agreement  with experiment. While this is the case,
it turns out that this ratio is very sensitive to the pair of values
$\beta_1 - \beta_2$, with a strong correlation between them.
In Fig. 4 we show the theoretical prediction of
the ratio $\mu_p G_E(q^2)/G_M(q^2) $ corresponding to
$\beta_1 = 3.0$ and $\beta_2 = 4.2$, together with the JLab data
\cite{RATIO2}.  Small variations of these parameters lead to correlated 
pairs resulting in equally good fits, e.g.
the pair $\beta_1 = 2.95$ and $\beta_2 = 4.13$
leads to an almost identical theoretical prediction. Exploring the
$\beta_1$-$\beta_2$ parameter space, and performing 
a combined fit to $F_1(q^2)$, $F_2(q^2)$, and the ratio
$\mu_p G_E(q^2)/G_M(q^2) $ gives
%Eq.9
\begin{eqnarray}
\begin{array}{lcl}
\beta_1 &=& 2.95 - 3.03 \\
\beta_2 &= & 4.13 - 4.20 \;.
\end{array}
\end{eqnarray}

The mean-squared electromagnetic radii that follow from Eq.(7) are given by
%Eq.10
\begin{equation}
<r^2_{1,2}> = 6\; \alpha'\; [\psi(\beta_{1,2} - 1/2) - \psi(1/2)] \; ,
\end{equation}
where $\psi(x)$ is the digamma function. Using the results from Eq.(10)
in Eq. (11) gives
$<r^2_1>^{1/2} = 0.72\; fm$, and $<r^2_2>^{1/2} = 0.78\; fm$. From Eq.(2), these
radii lead to the Sachs radii $<r^2_E>^{1/2} = 0.81 \; fm$, and
$<r^2_M>^{1/2} = 0.76 \; fm$. These values are in reasonable agreement with
various other results in the literature \cite{MERGELL}, especially taking into
account that the free parameters $\beta_{1,2}$ have been fixed from the large-$Q^2$
data, while the form factors decrease by 3-4 orders of magnitude in the
range $0 \leq Q^2 \simlt \; 30\; \mbox{GeV}^2$. It should also be kept in mind
that the strong deviation from unity of the ratio $\mu_p G_E(q^2)/G_M(q^2)$
might affect existing extractions of the Sachs radii from data.

In summary, the nucleon form factors $F_1(q^2)$, and $F_2(q^2)$ obtained in
the framework of Dual-$\mbox{QCD}_{\infty}$ reproduce very nicely the
experimental data in the space-like region, as corrected in \cite{BRASH},
in the wide range $0 \leq Q^2 \simlt \; 30\; \mbox{GeV}^2$.  
The Sachs magnetic form factor, $G_M(q^2)$, as well as the non-trivial
ratio $\mu_p G_E(q^2)/G_M(q^2)$ can also be accounted for in this framework.
These results provide strong support for Dual-$\mbox{QCD}_{\infty}$ as
a viable realization of QCD in the large $N_c$ limit.

Acknowledgements\\
One of us (CAD) wishes to thank F. Maas for a very helpful discussion on the experimental data.\\

\begin{center}
{\bf Figure Captions}
\end{center}

Figure 1. Dual-$QCD_\infty$  form factor $F_1(Q^2)$, Eq.(7),
for the fitted parameter $\beta_1 = 3.03$, together
with the experimental data as corrected in \cite{BRASH}.

Figure 2. Dual-$QCD_\infty$  form factor $F_2(Q^2)$, Eq.(7),
for the fitted parameter $\beta_2 = 4.20$, together
with the experimental data as corrected in \cite{BRASH}.

Figure 3. Dual-$QCD_\infty$  form factor $G_M(Q^2)$, Eq.(2),
for the fitted parameters $\beta_1 = 3.03$, and 
$\beta_2= 4.20$, together
with the experimental data as corrected in \cite{BRASH}.

Figure 4. Dual-$QCD_\infty$  ratio  $\mu_p G_E(Q^2)/G_M(Q^2)$
for the fitted parameters $\beta_1 = 3.00$, and 
$\beta_2= 4.20$, together
with the experimental data  \cite{RATIO2}.

\newpage
\begin{figure}[tp]
\epsffile{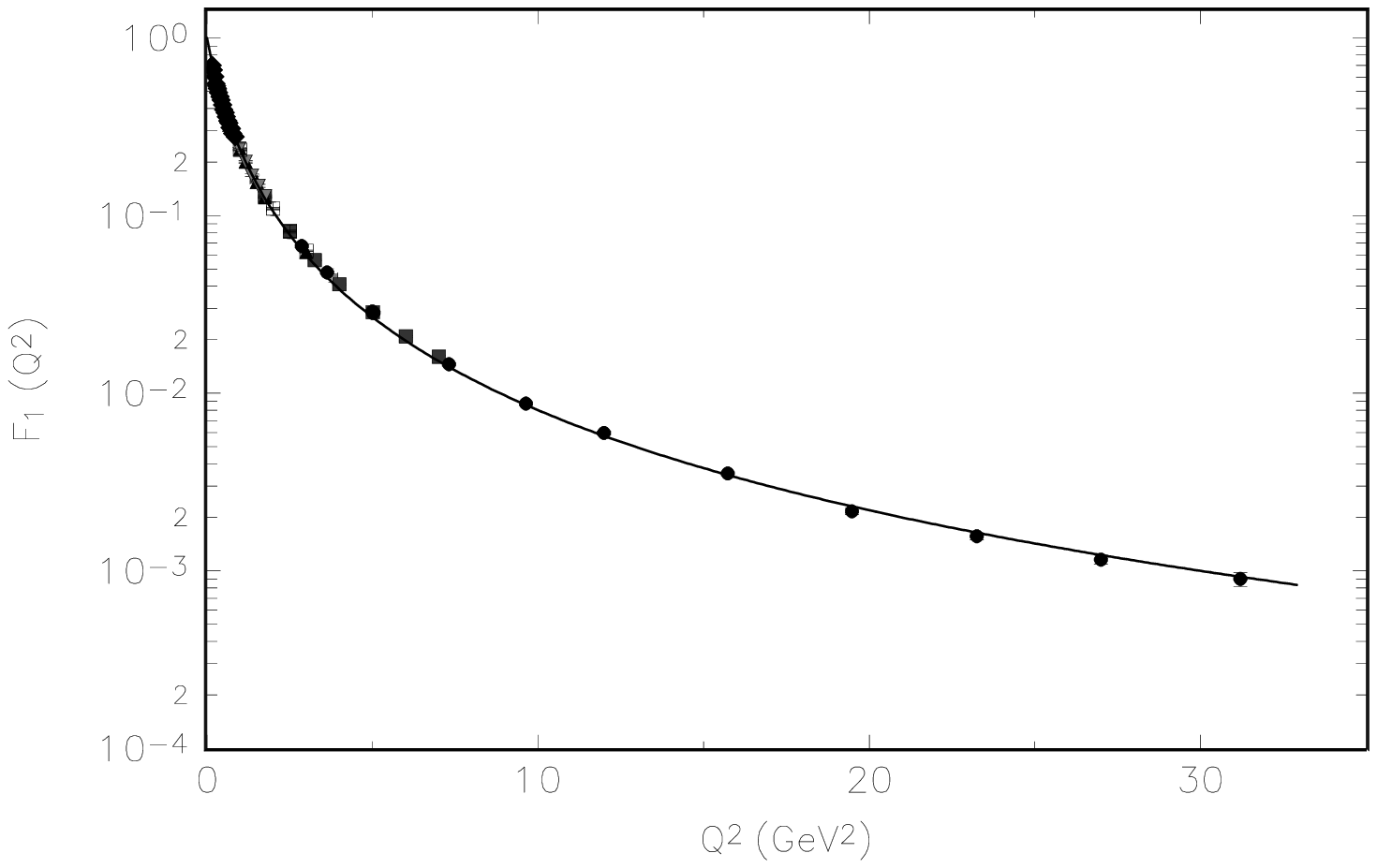}
\caption{}
\end{figure}
\newpage
\begin{figure}[tp]
\epsffile{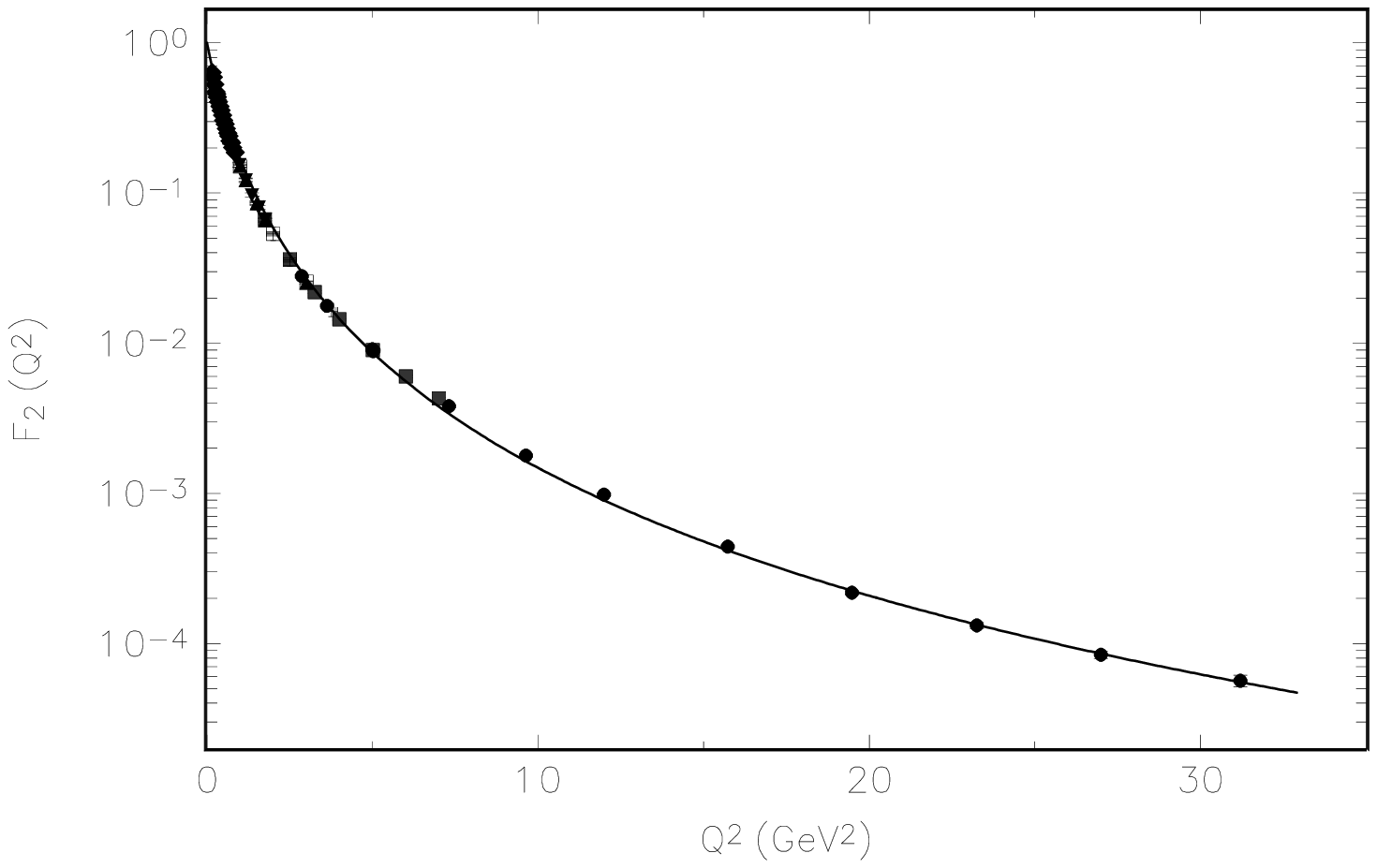}
\caption{}
\end{figure}%
\begin{figure}[tp]
\epsffile{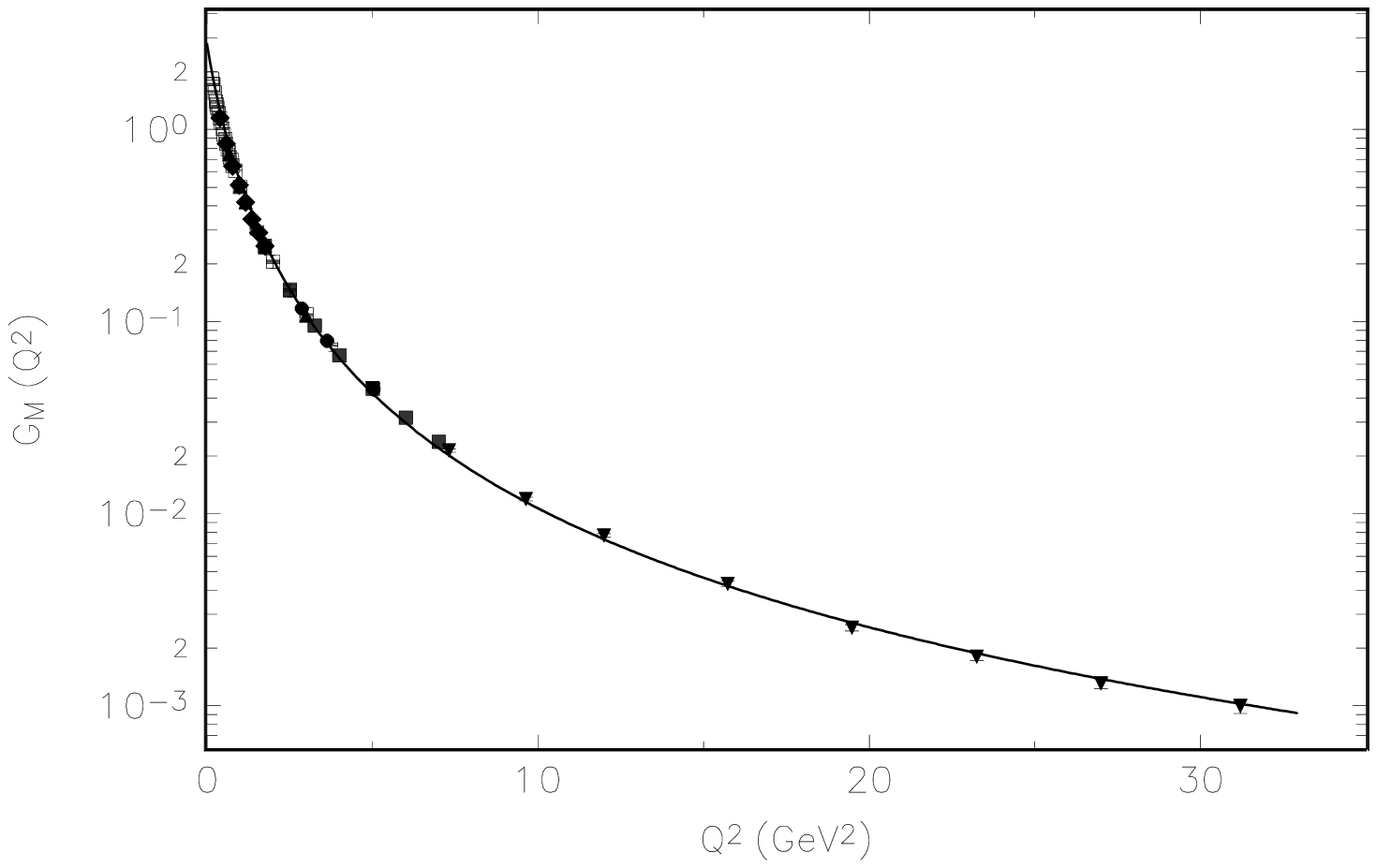}
\caption{}
\end{figure}
\newpage
\begin{figure}[tp]
\epsffile{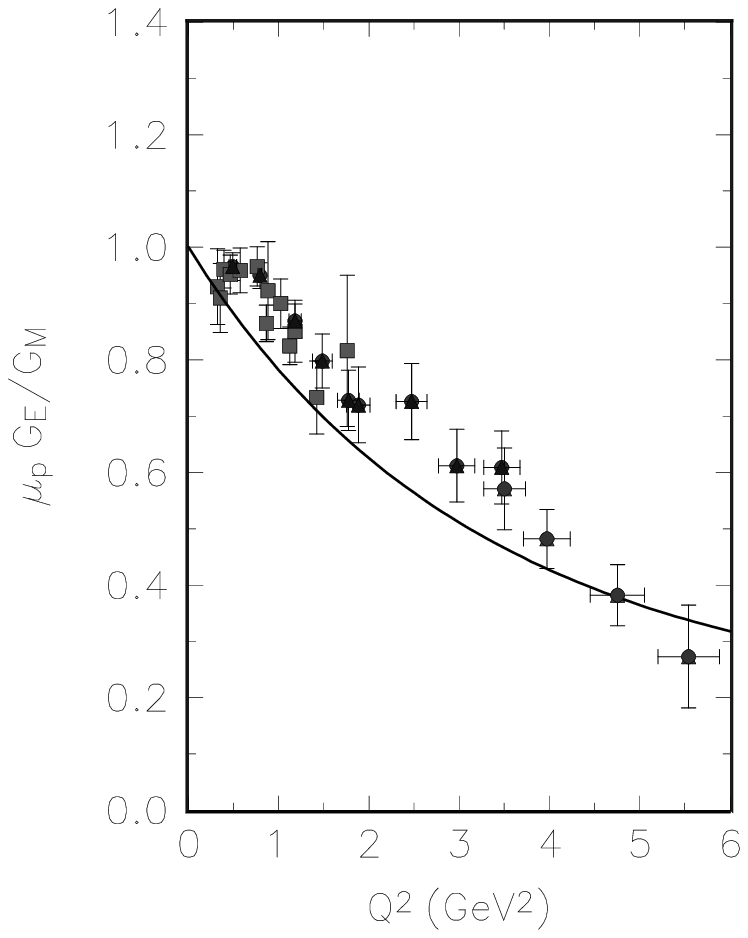}
\caption{}
\end{figure}

\end{document}